# Paradox in Wave-Particle Duality[*]

**Shahriar S. Afshar, Eduardo Flores, Keith F. McDonald, and Ernst Knoesel**

Department of Physics & Astronomy, Rowan University, Glassboro, NJ 08028;

afshar@rowan.edu

> *"All these fifty years of conscious brooding have brought me no nearer to the answer to the question, 'What are light quanta?' Nowadays every Tom, Dick and Harry thinks he knows it, but he is mistaken."* (Albert Einstein, 1951)

We report on the simultaneous determination of complementary wave and particle aspects of light in a double-slit type *welcher-weg* experiment beyond the limitations set by Bohr's Principle of Complementarity. Applying classical logic, we verify the presence of sharp interference in the single photon regime, while reliably maintaining the information about the particular pinhole through which each individual photon had passed. This experiment poses interesting questions on the validity of Complementarity in cases where measurements techniques that avoid Heisenberg's uncertainty principle and quantum entanglement are employed. We further argue that the application of *classical* concepts of waves and particles as embodied in Complementarity leads to a logical inconsistency in the interpretation of this experiment.

---

[*] A Preliminary version of this paper was presented by S.S.A. at a Seminar titled "Waving Copenhagen Good-bye: Were the Founders of Quantum Mechanics Wrong?," Department of Physics, Harvard University, Cambridge, MA 02138, 23 March, 2004.

# 1. INTRODUCTION

Wave-particle duality as embodied in Bohr's Principle of Complementarity (BPC) has been a cornerstone in the interpretation of quantum mechanics and quantum measurement theory since its inception.[1] The celebrated Bohr-Einstein debate[2] revolved around this issue and was the starting point for many of the illuminating experiments conducted during the past few decades. Bohr strongly advocated that "the unambiguous interpretation of any measurement must be essentially framed in terms of the *classical* physical theories".[3] This insistence on the primacy of classical concepts and logic in describing experiments led him to the introduction of his controversial principle of Complementarity as embodied in the following quotation: "…we are presented with a choice of *either* tracing the path of the particle, or observing interference effects… we have to do with a typical example of how the complementary phenomena appear under *mutually exclusive* experimental arrangements".[2]

In the context of the double-slit *welcher-weg* experiments, the original formulation of the BPC dictates that in a particular experimental configuration, "the observation of an interference pattern and the acquisition of which-way information are mutually exclusive".[4-9] Experiments have revealed the possibility of partial fringe visibility and partial which-way-information within strict limitations, and many experiments have backed this validation of BPC.[10-14] What these experiments have in common, however, is the fact that they provide information by measurement techniques which ultimately perturb the wavefunction.

In this paper we report on the presence of sharp interference and highly reliable which-way information in the same experimental arrangement for the *same* photons using *non-perturbative* measurement techniques at separate spacetime coordinates, both of which refer back to the behavior of the photon at the same event, i.e. the passage through the pinholes. We *inferred* full fringe visibility from the observation that the total photon flux was only slightly decreased when thin wires were placed exactly at the minima of the presumed interference pattern. Which-way information was obtained further downstream through the known imaging capabilities of a lens system. In the framework of classical logic, we make statements about the which-way information of the photon as it passes the plane of the pinholes. With respect to the mutual exclusivity of complementary wave and particle natures as expressed in BPC, the applied technique appears to allow us to circumvent the limitations imposed by Heisenberg's uncertainty principle and the entanglement between the which-way marker and the interfering quantum object as employed in some *welcher-weg* experiments.[14-17] Afshar's non-perturbative measurement technique[18-20] used in this work is conceptually different from quantum non-demolition[21] or non-destructive[22] techniques which do not destroy but perturb the photon wavefunction directly. The observation, that the presence of the wire grid decreases the photon count only negligibly, characterizes a confirmation null result. Such a null result represents a confirmation measurement in quantum mechanics, which verifies the expectation of a vanishing wavefunction at particular positions, in this case at the wire grid.

## 2. EXPERIMENT

The present experiment is an improved version of an experiment first suggested and carried out at high photon flux by Afshar.[18] After passing through a small aperture, light from a diode laser (wavelength = 638 nm) was incident onto a pair of pinholes with hole diameters of 40 µm and center-to-center separation of 250 µm. The two emerging beams from the pinholes, which were the sole light sources in this experiment, spatially overlapped in the far-field and interfered to produce a pattern of alternating light and dark fringes. At a distance of 0.55 m from the pinholes six thin wires of 127 µm diameter were placed at the minima of the interference pattern with an accuracy of ±10 µm. The wire-to-wire separation was ~1.3 mm. When the interference pattern was present the disturbance to the incoming beam was minimal. However, when the interference pattern was not present the wire grid obstructed the beam and produced scattering, thus reducing the total flux at the image plane of the detectors. From comparative measurements of the total flux with and without the wire grid we *inferred* the presence of an interference pattern in a non-perturbative manner.

To obtain the which-way information, the dual pinhole was imaged by a lens system with a magnification of ~ 4× onto two single-photon detectors (Perkin Elmer, SPCM-AQR series), which recorded the integrated flux at the image of the two pinholes. When the wire grid is not present, quantum mechanics predicts that a photon that hits detector 1 (2) originates from pinhole A (B) with a very high probability due to the one-to-one relationship between the pinholes and the corresponding images. Such application of an imaging lens for obtaining which-way information in the double-slit type experiments has been previously discussed in the literature,[23,24] however, it is not crucial to use a lens system for this purpose. An analogous experiment[19] was performed without the use of a lens, employing two coherent beams that intersected at a small angle. The wire was placed at the intersection of the beams where dark fringes were expected. In the far field beyond the region of overlap, the beams maintained their which-way information due to the law of conservation of linear momentum.[19, 20]

The current experiment was conducted in the low-flux regime to preclude loss of which-way information due to the intrinsic indistinguishability of coherent multi-photon systems. To illustrate this point, consider two correlated photons crossing an opaque screen having pinholes A and B. Assume that one photon passes pinhole A in state $|1,A\rangle$, and the other passes pinhole B in state $|2,B\rangle$. Due to the bosonic nature of the photon, the state of the two-photon system is the symmetric combination $\frac{1}{\sqrt{2}}[|1,A\rangle|2,B\rangle+|1,B\rangle|2,A\rangle]$. Accordingly, it is impossible to tell which photon passes through which pinhole. Each photon has a non-vanishing value at both pinholes. In this case the which-way information is fully lost. To avoid this problem we used a continuous wave laser at low photon flux. When the flux was $3\times10^4$ photons/sec, the average separation between successive photons was about 10 km, which was much greater than the coherence length (0.4 m) of the laser. Therefore, the probability of two photons passing through the experimental setup within the coherence length was very small. Experimentally, we determined the rates of coinciding photons (within 20 ns) on different detectors to be <1:10000 for a flux of ~ $3 \times 10^4$ photons/sec, and 5:2000 for a flux of ~ $10^7$ photons/sec. We thus conclude that the observed rate of coincidence was of statistical origin and too small to influence our results.

## 3. RESULTS

The intensity profiles for four different experimental conditions are shown in Fig 1. In case (a), when the wire grid was removed and both pinholes were open, we observed sharp images of the two pinholes. When the wire grid was properly positioned at the interference minima and both holes were open, case (b), only a slight reduction in peak intensity was observed, and there was no evidence of diffraction by the wires. However, when one of the holes was blocked and the wire grid was inserted, cases (c) and (d), diffraction patterns arose and the peak intensity was reduced drastically.

To quantify this observation we recorded the integrated counts of the image of each pinhole with two single-photon detectors at a flux of $3 \times 10^4$ photons/sec. With the wire grid removed and both pinholes open, case (a), the photon counts at detectors 1 and 2 were almost equal and above $10^6$. When we blocked one pinhole we found that the photon count at the corresponding detector was reduced to the dark-count level while the photon count at the other detector remained unaltered. Based on the known imaging capacity of the lenses,[23,24] we conclude that we had full which-way information when one or two pinholes were open and the wire grid was not present.

When the wire grid is positioned in the path of the beam and one of the pinholes is closed, cases (c) and (d), we expect a certain fraction of the photons to be scattered and absorbed by the wires. In fact, we observed a 14.14% reduction in the photon count at detector 1 if only pinhole A was open and 14.62% at detector 2 if only pinhole B was open. In contrast, when two pinholes were open and the wire grid was in place, case (b), the photon count for detector 1 decreased by only 0.31% from the case without the wire grid. The photon count for detector 2 decreased by 1.13% from the case without wire grid. These losses were due to the finite thicknesses of the wires and imperfect alignment. We conclude from the data above and from the absence of any substantial diffraction pattern (see Fig. 1b) that, at least, the *destructive* two-hole interference pattern was fully developed at the position of the wires.

## 4. ANALYSIS

### 4.1 Fringe Visibility

A thin wire grid is well suited for the determination of the fringe visibility when interference fringes are present. To illustrate this point, let us consider the ideal two-pinhole interference pattern described by $I = I_o \left( \frac{2J_1(au)}{au} \right)^2 \cos^2(bx)$, where $J_1$ is the Bessel function of order 1, $u$ is the radial distance from the center of the pattern, $x$ is the position along the horizontal axis, and $a$ and $b$ are constants.[25] Near the center of the interference pattern the term that contains the Bessel function is nearly 1, and the $\cos^2(bx)$ term is the dominant factor in the formula. By expanding the $\cos^2(bx)$ term the irradiance near a dark fringe is given by $I = I_o b^2 s^2$, where $s$ is the distance from the center of the dark fringe. Thus, if a wire of thickness $t$ were placed at the center of a dark fringe the maximum irradiance on a wire would be $I_o b^2 (t/2)^2$. Using $I_o$ for $I_{Max}$, and $I_o b^2 (t/2)^2$ for $I_{Min}$ in the standard formula for visibility,[25] $V = \dfrac{I_{Max} - I_{Min}}{I_{Max} + I_{Min}}$, we derive $V = 1 - \dfrac{1}{2} b^2 t^2$. In our experiment $b = 2.462/\text{mm}$ and $t = .127\,\text{mm}$, which results in a theoretical prediction for the visibility $V = 0.95$ for our setup, assuming that the interference pattern is the ideal two-pinhole one. It is important to notice that as the wires get thinner ($t \to 0$), the visibility approaches 1.

The very small decrease in the photon count (0.31% and 1.13% for each image respectively), when the wire grid was in place, is a strong evidence for the presence of a nearly perfect two-pinhole interference pattern with theoretical visibility $V \sim 0.95$. Obviously, we cannot measure the visibility of the pattern directly without compromising the which-way information. However, we *can* provide a lower limit compatible with our data. We assume an interference pattern with the worst possible visibility, made up of a periodic square function, which has a flat envelope across the Airy disk. Each bar of low irradiance has a width equal to the thickness of the wires and is assumed to be located at the positions of the wires. Thus they cumulatively cover an area equal to the total cross section of the six wires within the Airy disk, which can be approximated by $A_W \leq 6 \times 2Rt$, with $R = 10.7$ mm as the radius of the Airy disk and the wire thickness $t$. Similarly, the high irradiance bars have a net area equal to the area of the central maxima ($\pi R^2$) minus the area of the wires. From the data of figure 1, we conclude that the fraction of photons that are stopped by the wire grid is maximally 1%. The fraction of photons that pass the wire grid is >99%. The irradiance $I$ for the high and low bars of the periodic square function, $I_{Max}$ and $I_{Min}$, is now directly proportional to the fraction of photons divided by the respective area. A simple calculation gives a ratio of $I_{Max}/I_{Min}$ = 4.7. Using this value in the standard formula for $V$ we obtain the *lowest* possible limit in the worst case scenario for the visibility of any interference pattern compatible with our data of $V \geq 0.64$.

**4.2 Which-way Information**

Using classical logic, we define the partial which-way information in this experiment $K_B$ as the probability that a photon passing through pinhole B will hit the corresponding detector 2. In our experiment we measure a normalized photon count at detector 2, $W_2 = W_{A2} + W_{B2} = 98.87\%$, which is composed of photons originating from pinhole, B, $W_{B2}$ and to a smaller extent from photons originating from pinhole A, $W_{A2}$. The which-way information for photons from pinhole B is given by

$$K_B = \frac{W_{B2} - W_{A2}}{W_{A2} + W_{B2}} = 1 - \frac{2W_{A2}}{W_2}. \tag{1}$$

According to the principle of superposition, the electric field at detector 2, $E_2$, is given by the sum of the electric fields when one pinhole at a time is blocked: $E_2 = E_2^{A\,blocked} + E_2^{B\,blocked}$. Assuming constructive interference between the E-fields we derive

$$W_2 = E_2^2 = \left(E_2^{A\,blocked}\right)^2 + 2 E_2^{A\,blocked} \cdot E_2^{B\,blocked} + \left(E_2^{B\,blocked}\right)^2. \tag{2}$$

From the total photon count, $W_2$, we need to extract the contributions that come from pinhole A. The first contribution is $\left(E_2^{B\,blocked}\right)^2 \sim 0.46\%$, which relates to the photons from pinhole A scattered by the wires onto detector 2 when pinhole B is blocked (see data in figure 1d). The second contribution is the fraction of the photon count of the cross term, $2E_2^{A\,blocked} \cdot E_2^{B\,blocked}$, which stem from pinhole A. We estimate this fraction according to the ratio of the respective electric field $E_2^{B\,blocked}/E_2$. After adding the two contributions it can be shown that the percentage of photons from pinhole A at detector 2 has an upper limit of $W_{A2} \leq 3\left(E_2^{B\,blocked}\right)^2$. The partial which-way information is then given by $K_B \geq 1 - 6\left(E_2^{B\,blocked}\right)^2/E_2^2$, and we can calculate a value for the which-way information of $K_B \geq 0.97$ for our experimental data. Since the pinholes are the only light sources having equal flux, we conclude that when a photon is measured at detector 2 it must have originated from pinhole B with a probability of $K_B$ or higher. Similarly, for photons impinging on detector 1 we calculate the same lower limit of which-way information of $K_A \geq 0.97$.

## 5. DISCUSSION AND INTERPRETATION

At this point we wish to discuss the validity of our measurements with regard to the requirements that $K$ and $V$ have to be measured within the same experimental setup. We emphasize that the which-way information $K$ and the visibility $V$ are defined for the same experimental configuration, i.e. when the wires are positioned at the center of the minima of the presumed interference fringes, and both pinholes are open. To obtain values for $K$ and $V$ we perform three measurements of the photon count for the following three distinct configurations: i) no wire grid, ii) wire grid in central minima, iii) wire grid in central minima and one pinhole blocked. These measurements only serve to derive numerical values for $K$ and $V$ with the application of the superposition principle. They do not affect the complementary wave-particle aspect of the measured photons. Therefore, we argue that the above-described methods are valid in the derivation of $K$ and $V$.

Since the measurements of $K$ and $V$ are performed at two different places and at different times, one may argue that the wave and particle aspects are not present *simultaneously*. However, we can readily respond to this criticism by pointing out that both of the complementary measurements refer back to what "takes place" at the pinholes when a photon passes that plane. The which-way information tells us through which pinhole the particle-like photon had passed with $K \geq 0.97$. The interference indicates that the wave-like photon must have sampled both pinholes so that an interference pattern with $V \geq 0.64$ could be formed. Thus, the derived values for $K$ and $V$ of the photon refer back to the same space-time event, i.e. to the moment when the single photon passed the plane of the pinholes.

When the value of $K \geq 0.97$ is combined with our previous result for the visibility of the interference pattern, $V \geq 0.64$, we get $V^2 + K^2 \geq 1.35$, so that the Greenberger-Yasin inequality $V^2 + K^2 \leq 1$ appears to have been violated in this setup. This inequality, which can be derived from Heisenberg's uncertainty principle in the case of *perturbative* measurements, has been verified in numerous related *welcher-weg* experiments. However, in most of these experiments, the perturbative measurement of $K$ occurred *first*, by either tagging the quanta by internal state markers such as polarization, or by entangling them with another system. The perturbed photons then interfered and formed a pattern from which the visibility $V$ is derived. The perturbation caused by the measurement of $K$ diminishes the fringe visibility, and as expected, $K$ and $V$ obey the Greenberger-Yasin inequality in such experiments. In contrast, in our setup, *first* the visibility was determined with high accuracy and minimal wavefunction perturbation, and only *afterwards* was the which-way information obtained by an imaging lens in a destructive measurement process.

It is also important to realize that our measurement of the visibility is *essentially* different from the common procedure, because we use a mainly non-perturbative measurement technique.[18-20] While the minimum irradiance $I_{Min}$ was measured in a destructive manner, i.e. real loss of photons at the wires, the maximum irradiance $I_{Max}$ was inferred from a model assuming the worst-case scenario for an interference pattern. Thus, no direct measurement of $I_{Max}$ has been conducted and no photons have been destroyed in this process. Therefore, these *same* photons contributed to the determination of the which-way information, $K$, further downstream. It is noteworthy to emphasize that in the case of diminishing wire thickness, the perturbation of the beam becomes even smaller, thus increasing the reliability of the which-way information, while surprisingly, the visibility *also* approaches unity. To resolve this apparent contradiction with the Greenberger-Yasin inequality, we define $V^*$ as a quantity, which is derived from a *non-perturbative* measurement process discussed earlier. For this type of measurement the Heisenberg uncertainty principle is circumvented, thus one could argue that the Greenberger-Yasin inequality is not applicable.

While it is possible to reason the violation of the Greenberger-Yasin inequality, the situation is more subtle when we aim to interpret the definition of $V^*$ in terms of the wave nature of the photon. On the one hand, the Greenberger-Yasin inequality corresponds to a statement of BCP which includes direct measurements of partial fringe visibility and partial which-way information. Therefore, one could potentially argue that BCP may not apply to our experiment because we used a non-perturbative measurement technique. On the other hand, while the values of $V$ and $K$ are strongly associated with wave and particle nature of quanta, it is not clear how we can interpret the values of $V^*$ in the same context, although it is inconceivable to us as to how one could argue against the reality of the destructive interference at the wires. Thus the results of this experiment leave us with an unresolved paradox regarding the scope of classical language of waves and particles when applied to quantum mechanical systems.

## 6. CONCLUSION

In light of the experimental evidence presented in this paper, should we insist on the use of the classical language employed by Bohr, we would be forced to agree with Einstein's argument against Complementarity as eloquently expressed by Wheeler that "… for quantum theory to say in one breath 'through which slit' and in another 'through both' is logically inconsistent".[26] We look forward to a lively debate on the role of non-perturbative techniques in quantum measurement and its application to interpretations of quantum mechanics.

# AKNOWLEDGMENT

S.S.A. wishes to thank D.W. Glazer, G.B. Davis, J. Grantham and other supporters of this project. E.F. and E.K. would like to thank B. Kerr, Camden County College, and T.F. Heinz, Columbia University, for their support.

**FIGURE CAPTION**

Fig. 1: Intensity profiles for four different experimental conditions at a flux of $3 \times 10^7$ photons/s presented on a logarithmic scale. (a) Both pinholes open without wire grid, and (b) both pinholes open with wire grid, are very similar, and the peak intensity in (b) is only slightly reduced to 98%. When only one pinhole is open, (c) and (d), the peak intensity at the respective detector drops to 85%. In addition, the wire grid creates a diffraction pattern and a small photon count is measured at the other detector, i.e. 0.46% when pinhole B blocked (d), and 0.41% when pinhole A blocked (c).

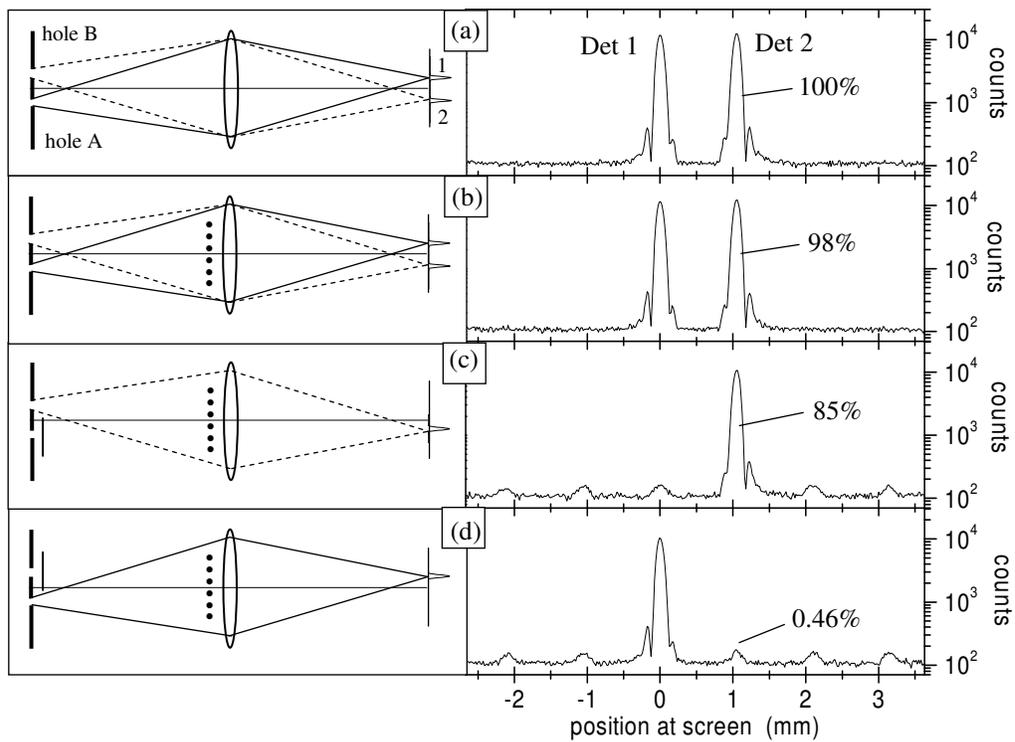

Figure 1